\documentclass[iop,apjl]{emulateapj}   

\pdfoutput=1

\usepackage{graphicx}
\usepackage{epsfig}
\usepackage{amsmath}
\usepackage{subfigure} 
\usepackage{hyperref}
\usepackage{amssymb}
\usepackage{apjfonts}
\usepackage{latexsym}
\usepackage{color}
\usepackage[usenames,dvipsnames,svgnames,table]{xcolor}
\usepackage{hyperref}
\usepackage[all]{hypcap}    
\hypersetup{
    colorlinks,
    citecolor=blue,
    filecolor=blue,
    linkcolor=blue,
    urlcolor=blue
}

\def\kms{\ifmmode{~{\rm km~s^{-1}}}\else{~km s$^{-1}$}\fi}
\def\cc{\ifmmode{~{\rm cm^{-3}}}\else{~cm$^{-3}$}\fi}
\def\fesc{\ifmmode{{f_{esc}}}\else{$f_{\rm esc}$}\fi}
\def\fstar{\ifmmode{{f_\star}}\else{$f_\star$}\fi}
\def\lsim{\lower0.3em\hbox{$\,\buildrel <\over\sim\,$}}
\def\gsim{\lower0.3em\hbox{$\,\buildrel >\over\sim\,$}}

\def\enzo{{\sc Enzo}}
\def\moray{{\sc Moray}}
\def\yt{{\sc yt}}
\def\Ms{\ifmmode{~{\rm M_\odot}}\else{M$_\odot$}\fi}
\def\Zs{\ifmmode{~{\rm Z_\odot}}\else{Z$_\odot$}\fi}
\def\h2{H$_2$}

\def\nat{Nature}
\def\apj{ApJ}

\def\apjl{ApJL}
\def\apjs{ApJS}
\def\mnras{MNRAS}
\def\araa{ARA\&A}
\def\aap{A\&A}

\def\physrep{Physics Reports}

\shorttitle{X-ray Background from Pop III Remnants} 
\shortauthors{H. Xu et al.}

\begin{document}\title{X-ray Background at High Redshifts from Pop III Remnants: Results from Pop III star formation rates in the Renaissance Simulations}

\author{
  Hao Xu\altaffilmark{1},
  Kyungjin Ahn\altaffilmark{2},
  Michael L. Norman\altaffilmark{1},
  John H. Wise\altaffilmark{3}, and
  Brian W. O'Shea\altaffilmark{4}}

\affil{$^{1}${San Diego Supercomputer Center, University of California, San Diego, 9500 Gilman
    Drive, La Jolla, CA 92093;
    \href{mailto:hxu@ucsd.edu}{hxu@ucsd.edu},
    \href{mailto:mlnorman@ucsd.edu}{mlnorman@ucsd.edu}}}

\affil{$^{2}${Department of Earth Science Education, Chosun University,
Gwangju 501-759, Korea; \href{mailto:kjahn@chosun.ac.kr}{kjahn@chosun.ac.kr}}}

\affil{$^{3}${Center for Relativistic Astrophysics, School of Physics,
    Georgia Institute of Technology, 837 State Street, Atlanta, GA
    30332; \href{mailto:jwise@gatech.edu}{jwise@gatech.edu}}}

\affil{$^{4}${Department of Computational Mathematics, Science and
  Engineering, Department of Physics and Astronomy, and National
  Superconducting Cyclotron Laboratory, Michigan State University,
East Lansing, MI 48824; \href{mailto:oshea@msu.edu}{oshea@msu.edu}}}

\label{firstpage}

\begin{abstract}

Due to their long mean free paths, X-rays are expected to have global impacts on the properties of the intergalactic medium (IGM) by their large scale heating and ionizing processes. 
At high redshifts, X-rays from Population (Pop) III binaries might have important 
effects on cosmic reionization and the Ly$\alpha$ forest. 
As a continuation of our previous work
on Pop III binary X-rays \citep{Xu14}, we use the Pop III distribution 
and evolution from the \textit{Renaissance Simulations}, a suite of
self-consistent cosmological radiation hydrodynamics simulations of the
formation of the first galaxies, to calculate the X-ray luminosity density and background over the redshift range $20 \geq z \geq 7.6$.  As we find that Pop III star formation continues at a low, nearly constant rate to the end of reionization, X-rays are being continuously produced at significant rates compared to other possible X-ray sources, such as AGNs and normal X-ray binaries during the same period of time. 
We estimate that Pop III binaries produce approximately 6 eV of energy in the X-rays
per hydrogen atom. We calculate the X-ray background for different monochromatic photon 
energies. KeV X-rays redshift and accumulate to produce a strong X-ray background spectrum 
extending to roughly 500 eV.  The X-ray background is strong enough to heat the IGM to $\sim 1000$ K and to ionize a few percent of the neutral hydrogen.  These effects are important for an 
understanding of the neutral hydrogen hyperfine transition 21-cm line signatures, the Ly$\alpha$ 
forest, and the Thomson optical depth to the CMB. 

\end{abstract}


\section{Introduction}
\label{sec:introduction}

One of the most important problems in
astrophysics and cosmology in both theory and observation is to obtain
the thermal and ionization history of the intergalactic medium (IGM) prior, during, and after the epoch of reionization.
It is critical to understand cosmological observations, such as high redshift
 21-cm  neutral hydrogen line signatures, the Thomson scattering optical depth of the CMB, and
 properties of the Ly$\alpha$ forest.

The appearance of the first luminous objects---likely the first stars---marks the end of the cosmic dark ages after recombination, and the beginning of the last cosmic phase transition of reionization, a prolonged process that ionizes and heats the IGM. The first generation of stars (Population III stars) are believed to have
formed from metal-free gas in small dark matter halos and having a large (tens to  a few 
hundreds M$_{\odot}$) characteristic mass \citep{Abel02, Bromm02, OShea07a,
  Turk09, Greif12_P3Cluster, Hirano15, Hosokawa16}. Pop III stars have short
lifetimes \citep{Schaerer02} and may explode as supernovae (SNe), enriching their
surrounding IGM. Once the gas metallicity passes 
some critical threshold, $\sim$ 10$^{-6}$ Z$_{\odot}$ if dust 
cooling is efficient \citep{Omukai05,Schneider06, Clark08} or $\sim$ 10$^{-3.5}$
Z$_{\odot}$ otherwise \citep{Bromm01, Smith09}, the gas can cool
efficiently and lower its Jeans mass and form (Pop II) stars in clusters at much higher rates.  
This makes the Pop III period of a given galaxy short. Consequently it is believed that Pop III stars are not a major contributor to the UV photons responsible for reionization \citep[e.g.][]{Fan06}. (However see \citet{Wise08_Reion} and \citet{Ahn12} for their contribution to the {\em start} of reionization).

 X-ray radiation with its much longer mean free path 
 than UV radiation has been considered a good candidate for the pre-ionization and pre-heating 
 the IGM much earlier than the end of the reionization at $z \sim 6$ \citep[e.g.,][]{Oh01, Venkatesan01, Ricotti04, Ricotti05},
 and for smooth heating of the IGM; e.g. \citep{Haiman11}. 
Its impact on the high redshift 21cm signal has been explored by many authors \citep{Pritchard07,Mirabel11,PritchardLoeb12,Visbal12,Pacucci14,Fialkov14a,Mirocha16}, and observational constraints on the cosmic X-ray background have been investigated by \citet{Fialkov16}.
 Recent cosmological simulations \citep{Turk09, Stacy10, Susa14} have found that metal-free star-forming clouds might
fragment to form binaries in a non-negligible fraction of Pop III star
forming events, making Pop III stars and remnants possible strong X-ray sources 
at high redshifts. Pop III in the approximate mass range $40-140 \Ms$ and $>260 \Ms$ may 
directly collapse to form black holes \citep[BHs;][]{Heger03}. Any strong accretion onto
these massive Pop III seed BHs would lead to X-ray radiation at high
redshifts \citep{Kuhlen05, Alvarez09, Tanaka12, Visbal12}.  
X-rays from Pop III binaries have been suggested to produce a
pre-heated IGM \citep[e.g.,][] {Mirabel11, Haiman11, Visbal12, Mesinger13, Fialkov14b, Madau16},
and to partially ionize the IGM in large volumes \citep[e.g.,][]{Ostriker96, Pritchard07}. 
\citet{Xu14} showed that Pop III binaries might be the dominant X-ray sources at high redshifts and discussed how the IGM might 
be heated locally by hundred eV photons and globally by the X-ray background from keV photons. 
Homogeneously heating of the IGM on tens to hundreds Mpc 
scales by an X-ray background may be important to understand the linewidths of Ly$\alpha$ forest spectra  \citep{Tytler09} and 
high-z 21-cm  neutral hydrogen line signatures \citep{Ahn14}. 

Recently, \citet{Xu16_Late3} showed using numerical simulations that Pop III stars can form continuously at significant rates from redshifts higher 
than 20 down to the end of the reionization. This suggests that X-rays from Pop III binaries
might continue to be important X-ray sources till the end of the reionization and the X-ray background may heat the IGM
for a long enough period to significantly heat the IGM globally.  
In this {\em Letter}, we estimate the X-ray luminosities  from Pop III binaries using the results of the \textit{Renaissance 
Simulations} \citep{Xu13, OShea15, Xu16_fesc} and calculate the X-ray background using simple models.  This paper extends to lower redshifts the results first presented in \citet{Xu14} concerning the X-ray background itself; we defer to a future paper a detailed calculation of the heating and ionization of the IGM and its 21cm signature.  We first describe the simulations and 
X-ray source and background models in Section 2. In Section 3, we present our results for the X-ray luminosity density, and the
background intensity and spectra. We discuss the implications of this strong X-ray background from Pop III
stars in Section 4.

\section{The Renaissance Simulations and X-ray Models}
\label{sec:methods}

The \textit{Renaissance Simulations} are a suite of adaptive mesh refinement (AMR)
first galaxy formation simulations carried out using the \enzo\footnote{\url{http://enzo-project.org/}} code \citep{Bryan13} on the Blue Waters supercomputer operated by NCSA/UIUC.  
\enzo\ has been used extensively for high redshift structure formation research 
\citep[e.g.,][]{Abel02,OShea07a,Xu08b,Turk09,Wise12b,Wise12a,Xu13,Wise14}. 
Notably, UV ionizing radiation from stellar populations is followed 
using the \moray\ adaptive ray tracing radiation transport solver \citep{Wise11}. Lyman-Werner 
radiation transport from stellar sources is calculated without considering absorption
using ray-tracing as well. However, we included the attenuation of LW photons,
by a grey opacity approximation, the ``picket-fence" modulation factor, as a 
function of the comoving distance between the source and the observer \citep{Ahn09}. 
The baryonic properties are calculated using a 9-species primordial non-equilibrium 
gas chemistry and cooling network \citep{Abel97}, supplemented by metal-dependent
cooling tables \citep{Smith09}.  Prescriptions for
Population III and metal-enriched star formation and feedback are
employed which have been fully described in \citet{Xu13} and \citet{Wise12a}. Here we briefly summarize the Pop III prescriptions because they are the focus of this paper. Individual Pop III star particles are created when the overdensity, cooling time, $H_2$ fraction, and metallicity thresholds given in \citet{Wise12b} are obeyed. The stellar mass, which dictates its lifetime and fate is drawn from a top-heavy IMF with a charateristic mass of $40 M_{\odot}$.
The  UV hydrogen ionizing and LW photon luminosities and 
lifetimes of the Pop III stars are determined by the mass-dependent model 
from \citet{Schaerer02}. Depending on their mass, they can end their lives as Type II or pair instability SNe, or collapse directly to black holes \citep{Heger03}. Kinetic and metal yields for the SNe are taken from \citet{Heger03}. Pop III stellar remnants' positions are tracked for the duration of the simulation as they become incorporated into more massive halos. Their population statistics were examined in \citet{Xu13}. 

The simulated box is a region of the universe with 28.4 comoving Mpc/h on a side (or 40 Mpc with
$h=0.71$). This is more than adequate to the sample the large fluctuations in the space density of Pop III star-forming minihalos \citep{BarkanaLoeb2004}. We do not include the streaming velocity effect which modulates Pop III star formation on yet larger scales \citep{Fialkov13} because our Lyman-Werner regulated Pop III star formation occurs in minihalos with $M_h > 5 \times 10^6 M_{\odot}$ \citep{Xu13, Xu16_Late3} where the streaming effect is negligible \citep{Stacy11,Greif11}.
The simulations use the WMAP7 $\Lambda$CDM+SZ+LENS best fit cosmology \citep{Komatsu11}:
$\Omega_{M}=0.266$, $\Omega_{\Lambda} = 0.734$, $\Omega_{b}=0.0449$,
$h=0.71$, $\sigma_{8}=0.81$, and $n=0.963$.  Initial conditions were
generated at $z=99$ using MUSIC \citep{Hahn11_MUSIC}, and a
low-resolution ($512^3$ root grid) simulation was run to $z=6$ to find
regions suitable for re-simulation.  The simulation volume was then
smoothed on a physical scale of 6 comoving Mpc, and regions of high
($\langle\delta\rangle \equiv \langle\rho\rangle/(\Omega_M \rho_C) -1
\simeq 0.68$), average ($\langle\delta\rangle \simeq 0.09$), and low
($\langle\delta\rangle \simeq -0.26$) mean density (at $z=15, 12.5,
\mathrm{and}~8$, respectively) were chosen for re-simulation.  These
subvolumes, hereafter designated the ``\textit{Rarepeak,}''
``\textit{Normal},'' and ``\textit{Void}'' regions, with comoving
volumes of 133.6, 220.5, and 220.5 Mpc$^3$, were resimulated with
three additional static nested grids, resulting in an effective
initial resolution of $4096^3$ grid cells and particles in the region
of interest, translating to a dark matter mass resolution of $2.9
\times 10^4$~M$_\odot$ in the same region.  We allowed further
local refinement (and then, the star formation) in the Lagrangian volume of the finest nested grid based on
baryon or dark matter overdensity for up to 12 total levels of
refinement (i.e., a comoving resolution of 19 pc in the finest cells.)
For more details about the calculations and scientific results about \textit{Renaissance Simulations}, see
\citet{Xu13,Xu14,Chen14, OShea15}. 

The X-ray source and background models have been employed in \citet{Xu14} and \citet{Ahn14}. Here we summarize the basic assumptions and approach. The X-ray source model 
assumes that a fraction $\epsilon_{\rm bin} = 0.5$ of Pop III stars form in binaries, and evolve to become high mass X-ray binaries (XRB) which emit X-rays for a fixed lifetime $\tau$ = 30 Myr. We use the same luminosities as in \citet{Xu14}
to calculate the X-rays emitted in the AMR-refined volumes. To calculate the X-ray background, we need to estimate the X-ray emission coming
from the non-refined regions, in which star formation is not resolved. We do this in two different ways. In Method 1, we adopt the same simple prescription described in \citet{Xu14}; namely for every root grid cell in the refinement volume containing X-ray emitting sources, we sum up their baryon density $\rho$ and X-ray luminosity $L$. Dividing by the number of emitting cells we have their average values $\bar{\rho}$ and $\bar{L}$. Then, in the unrefined portion of the simulation volume, if the root grid cell's baryon density $\rho \geq \bar{\rho}$, we assign it an X-ray luminosity $\bar{L}$. The defect of this model is that it predicts different amounts of total emission from the unresolved region, which is mostly common to each simulation. This is due to the different mean densities of the refined regions.  To deal with this problem we introduce Method 2,  in which we combine the results from all three refined regions to predict the X-ray emission at every redshift. For each root grid cell in the three survey volumes at each redshift,
we record $L \geq 0, \rho$, and a smoothed density over a $\sim$ 1 Mpc sphere $\rho_s$ centered on the cell. From this data we apply machine learning algorithms from the Python toolkit {\tt scikit-learn} to derive two functions: $P_X(\rho, \rho_s, z)$ and $L_X(\rho, \rho_s, z)$ . $P_X$ is the normalized probability that a cell with values $(\rho, \rho_s, z)$ contains one or more X-ray sources. We assume the probability can be represented as a second order polynomial in $z, \rho$ and $\rho_s$
and fit it using all data from three simulations at all redshifts. $L_X$ is a second order polynomial fit to the dimensional X-ray luminosity for only those cells with $L > 0$. Armed with these two functions we now walk through the cells in the unrefined region and draw a random number to compare with $P_X$ to decide whether it contains X-ray sources.  
If it does, we assign it an X-ray luminosity using $L_X$.  In this way, we generate the X-ray emissivity field in the unrefined root grid for all redshift outputs in each of the three simulations.  
As we will see in the next section, Method 1 provides only a rough estimate, for the reason stated, and is not recommended. Method 2 is more robust because it is based on fitting far more data from different redshifts and environments. It is nonetheless worthwhile presenting both as it provides a quantitative estimate for the error.   

Given the X-ray emissivity everywhere in the 40 Mpc periodic volume, we perform the same calculations as in \citet{Xu14}, Sec. 5.2 to estimate the X-ray background. We locate the 3D field of X-ray 
luminosities frozen at the most recent past in the full box periodically, and calculate the contribution at every location
of the box by summing over the full contribution from all the X-ray sources but within the corresponding 
lookback time. We simply treat the background  sources to be uniformly distributed, since inhomogeneity in the 
source distribution at large lookback times will be observed to be almost uniform inside the box. Here, we 
assume that the whole simulation box is a good representation of the average universe and take its luminosity 
as the mean, globally averaged luminosity, and we also consider the absorption of photons by \ion{H}{1} and \ion{He}{1} with 
cosmic mean density.

 Simulation data were processed and analyzed using \yt\footnote{\url{http://yt-project.org}} \citep{yt_full_paper} on
 the Blue Waters system at NCSA, the Gordon system at SDSC, and the Wrangler system at TACC.

\section{Results}
\label{sec:results}

Before we show the results of X-ray generation and background, we illustrate the Pop III star formation 
rate densities from the three simulated volumes in Figure 1. The Pop III star formation rate density is simply correlated with the mean matter
overdensity. The \textit{Rarepeak} simulation has earlier Pop III star formation and higher formation rate. The \textit{Void} region is the only simulation 
we can run to single digit redshifts. Its Pop III star formation starts late at z $\sim$ 20 and the 
rate density increases to $\sim$ 10$^{-5}$ M$_{\odot}$ yr$^{-1}$ Mpc$^{-3}$ at z $\sim$ 13, and then gradually decreases 
below z $\sim$ 10. The Pop III star formation remains roughly constant for more than 400 Myr. The Pop III star formation of  the \textit{Normal} region also appears to plateau at z $\sim$ 13, albeit at a higher rate than the \textit{Void} region. As discussed in  \citet{Xu13,Xu16_Late3} the Pop III formation is regulated through suppression of their formation in low mass halos by the Lyman-Werner radiation from stellar feedback. However their formation is not suppressed in halos of mass $\sim 10^7 M_{\odot}$ and above, despite strong photodissociating radiation. The plateaus reflect the rates at which pristine halos of sufficient mass are forming in these different environments.

\begin{figure}
\includegraphics[width=1\columnwidth]{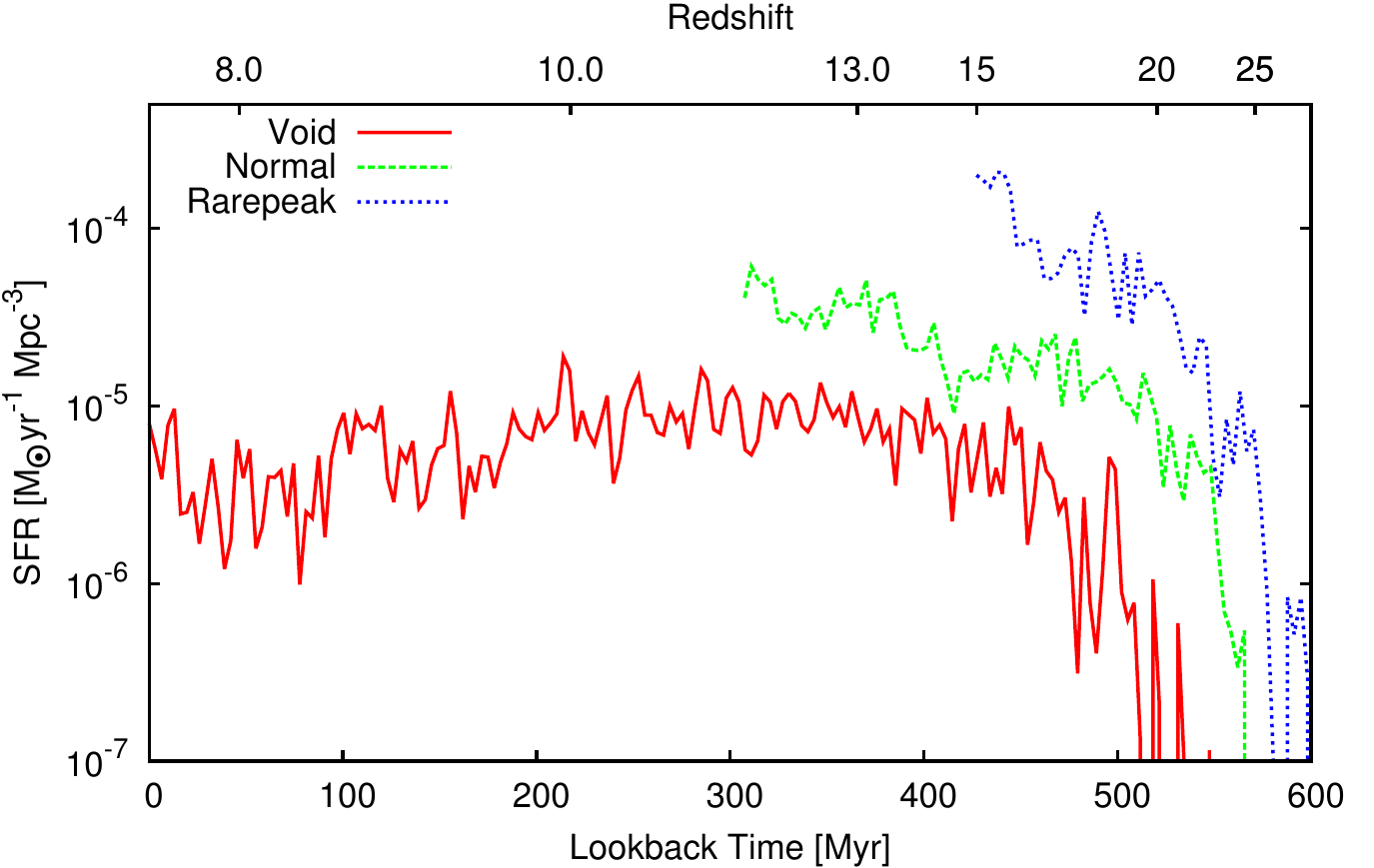}
\caption{Evolution of Pop III star formation rate densities of \textit{Rarepeak}, \textit{Normal} and \textit{Void} regions.
\label{fig:sfr}}
\end{figure}

Figure 2 shows the evolution of the X-ray luminosity densities of the refined regions and the total X-ray luminosities of the whole box. The X-ray
luminosity densities simply follow the Pop III star formation rate densities. Consequently, at $z=15$, the cosmic variance effect on the predicted X-ray emissivity density is about an order of magnitude.  Using Method 1,  the total X-ray luminosities are underestimated for the
\textit{Rarepeak}, and overestimated for the \textit{Void} run.  To better estimate the X-ray background, we scale down the total luminosity from the \textit{Void} region to 
match that from the \textit{Normal} simulation.  Using Method 2, we obtain a very smooth X-ray luminosity evolution. It grows from 1.7 $\times$ 10$^{43}$ to    
3 $\times$ 10$^{44}$ erg s$^{-1}$ from $z=20$ to $z=7.6$, as the luminosity density grows from 2.7 $\times$ 10$^{38}$ to 4.7 $\times$10$^{39}$ erg s$^{-1}$ Mpc$^{-3}$. 
This is a factor of 30($\epsilon_{\rm bin}/0.5)$ higher than estimates of the X-ray luminosities from normal X-ray binaries and AGN \citep{Fragos13, Madau16}. Taking the result from Method 2 as our reference, we would conclude that Method 1 is accurate to $\pm 50$\% at the redshifts where they overlap.

\begin{figure}
\includegraphics[width=1\columnwidth, clip=true]{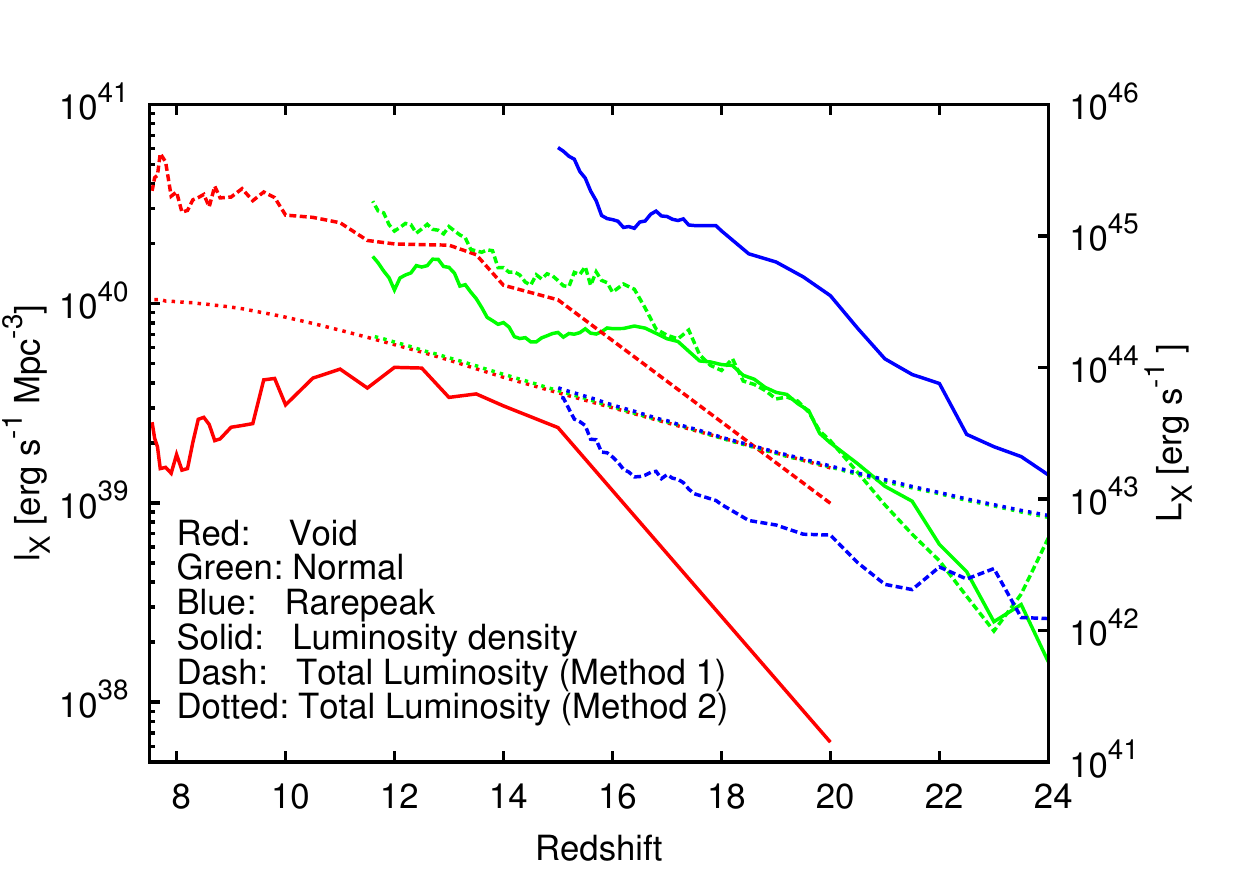}
\caption{Evolution of X-ray luminosity density in the refined regions (solid lines) and total X-ray luminosity of the entire (40 Mpc)$^3$ simulation volume (dashed and dotted lines).
Dashed lines show the total luminosity extrapolated from \textit{Rarepeak}, \textit{Normal}, and \textit{Void} regions, respectively,  using Method 1. Dotted
lines show the total X-ray luminosity obtained from Method 2. Since this method uses a single correlation function between X-ray emission and gas density 
and redshift, using the data from all three simulations together, the total X-ray luminosities predicted for the three runs are almost the same.  
\label{fig:xray_luminosity}}
\end{figure}

Mean background X-ray intensities are shown in Figure 3. We assume the X-ray sources are monochromatic and calculate the background for six different 
photon energies. The X-ray intensity at an observer redshift is the total intensity of observed redshifted X-rays from all earlier redshifts. In the upper panel, we 
show the X-ray background intensities from \textit{Void} and \textit{Normal} simulations using X-ray emissions calculated using Method 1. 
In the lower panel, we do the same thing using Method 2. In both cases the X-ray
background gradually builds up, and is significantly dependent on the emitted photon energy.  As discussed in \citet{Xu14}, only the high energy (keV) photons accumulate to form a global background since sub-keV photons are absorbed near the source. The X-ray
background intensity of 3 keV photons from Method 2, which we consider more reliable than Method 1, is over 10$^{-6}$ erg s$^{-1}$ cm$^{-2}$ sr$^{-1}$.

\begin{figure}
\includegraphics[width=0.9\columnwidth, clip=true]{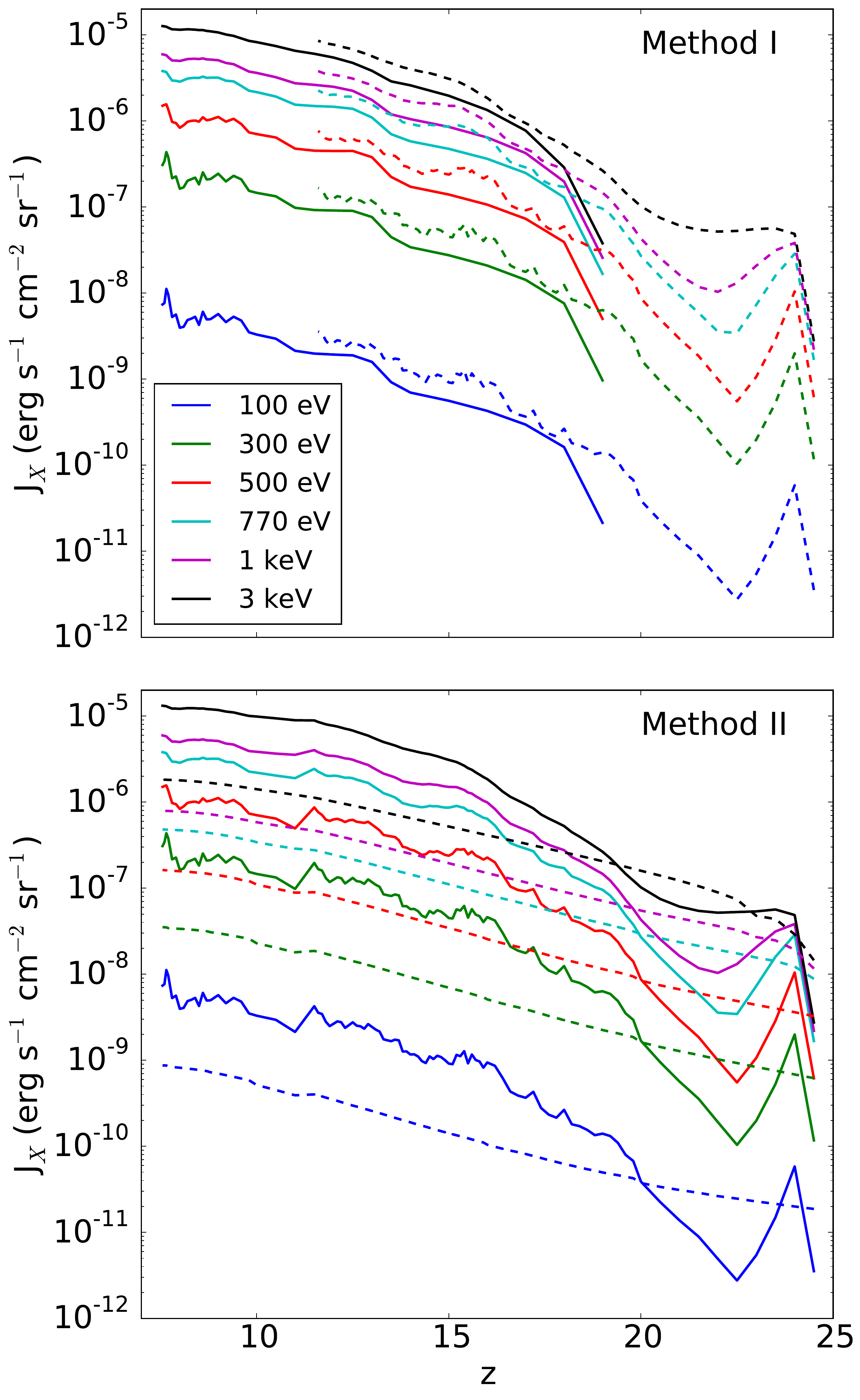}
\caption{Evolution of the mean X-ray background intensities for different observed X-ray energies. Upper panel: results from \textit{Void} (solid line) and \textit{Normal} (dashed line) simulations using Method 1. Lower
panel: same as upper panel but using Method 2. The background intensity is the summation of all the redshifted radiation observed.
\label{fig:intensity_evolution}}
\end{figure}

Figure 4 illustrates the X-ray background spectra for different source photon energies for different models at $z=12$ and $z=7.6$ showing how X-ray photons
with different energies from higher redshifts contribute to the X-ray background. Figure 5 shows the synthetic X-ray spectra at $z=12$ 
and $z=7.6$ by assuming the source X-ray energies are equally distributed over the six energy bands. The high energy keV photons with low absorption produce 
continuous broad band observed spectra by cosmological redshifting. Except for the gap between 1 and 1.5 keV for the $z=12$ spectrum due to the redshift cutoff, the X-ray spectra have 
similar strengths at $z=12$ and $z=7.6$, as the Pop III star formation rate does not evolve much. At $z=7.6$, keV X-rays from early times redshift to sub-keV energies and then 
heat and ionize the IGM. Using a one-zone calculation \citet{Xu14} showed that the X-rays of 1 keV and 3 keV in this spectrum have similar heating and ionizing effects to the IGM.
The efficient absorption of X-rays happens below $\sim$ 500-600 eV,  and any X-rays with photon energies lower than this are absorbed locally 
and only contribute a narrow band to the spectrum.   

\begin{figure}
\includegraphics[width=0.9\columnwidth]{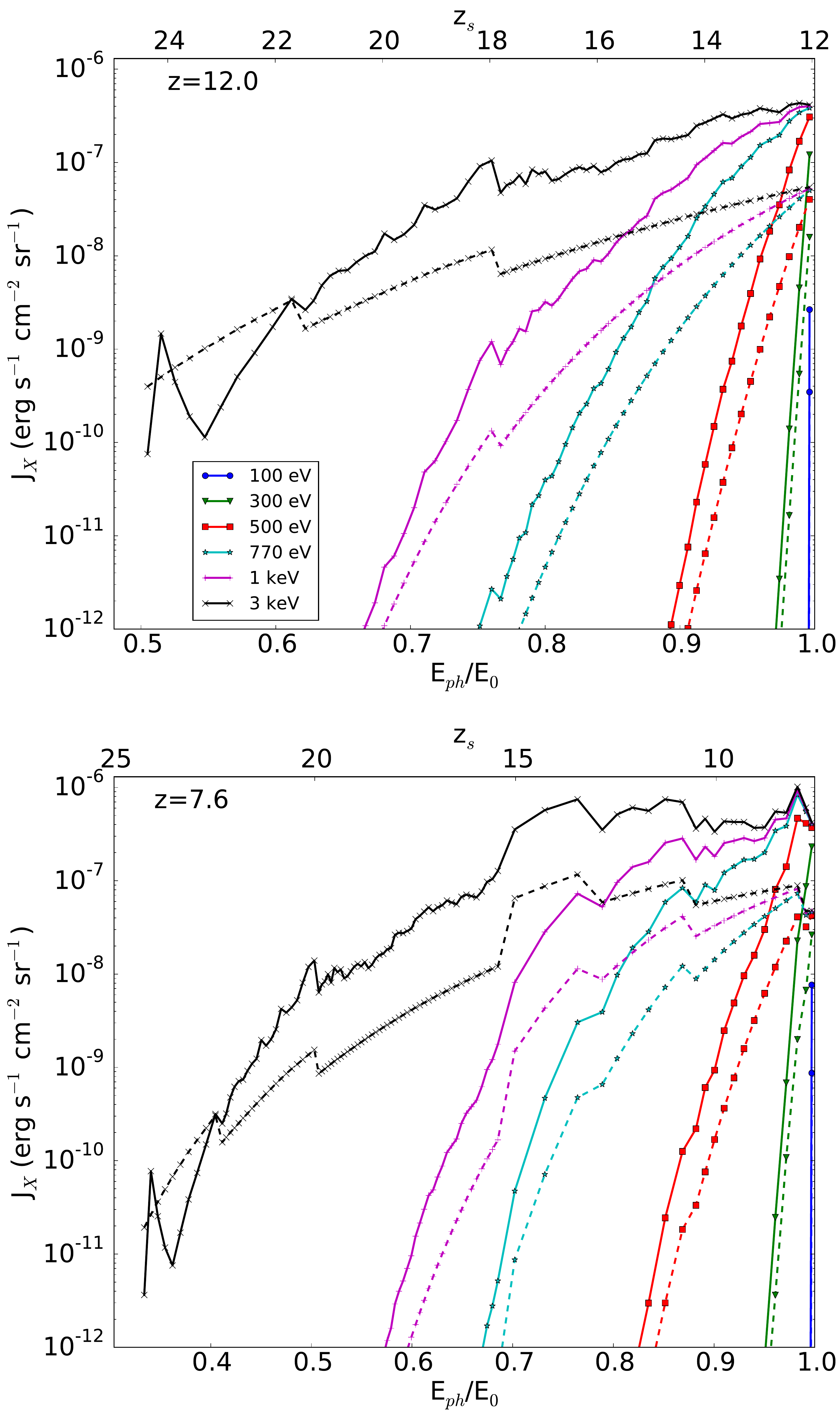}
\caption{Mean X-ray background spectra at redshifts $z=12$ and $z=7.6$ for different monochromatic X-ray source energies. Solid lines are for Method 1, and dashed lines are Method 2. The lower x-axis is the ratio of observed photon energy to source photon energy, while the 
upper x-axis is the source redshift.  
\label{fig:spectra}}
\end{figure}

\begin{figure}
\includegraphics[width=\columnwidth]{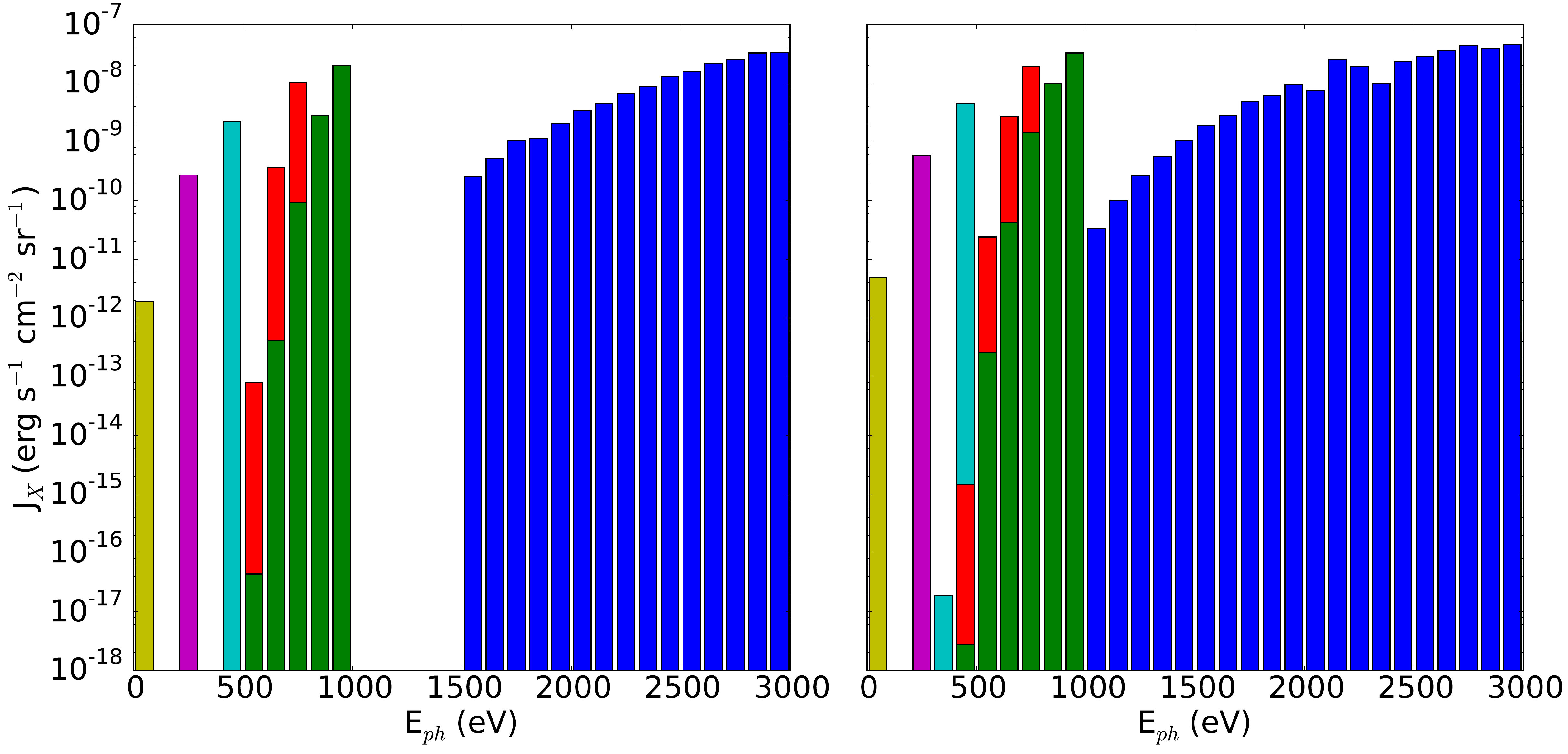}
\caption{ Synthetic X-ray spectra at $z=12$ (left) and  $z=7.6$ (right) from Method 2 by assuming source X-rays are equally distributed over all six energy bands. X-axis is the observed
photon energy. X-ray intensities of different energy sources are in different colors with the same scheme as Figure \ref{fig:spectra} and are stacked.  
\label{fig:synthetic_spectrum}}
\end{figure}

\section{Summary and Discussion}
\label{sec:conclusions}

In this {\em Letter}, we present estimates of the X-ray emission and background from Pop III binaries derived from the Pop III star formation histories
in \textit{Renaissance Simulations} beginning at $z \sim 25$ down to $z=7.6$ near the end of reionization. The key results are summarized as follows:

\begin{enumerate}

\item A significant amount of X-rays may be produced by Pop III binaries down to the end of reionization, as a direct result of
Pop III stars forming continuously with steady rates since they begin forming at redshifts of $20-30$.  

\item The X-ray background builds up gradually once the first stars begin forming. The X-ray background intensity depends on the assumed source
photon energies. Only keV photons contribute to the X-ray background, and become a broad band spectrum in the background.  

\item X-rays from Pop III binaries might continuously dominate X-rays from normal stellar X-ray
binaries and AGN until the end of reionization, although this conclusion depends on the assumed fraction of Pop III stars that become X-ray binaries. 

\end{enumerate}

The results of this {\em Letter} may have important implications on the understanding of the reionization process, the properties of the post-reionization IGM, and 
many cosmological observations, like the Thomson scattering optical depth of the cosmic microwave background, neutral hydrogen 21-cm 
transition line, and the Ly$\alpha$ forest.

The Pop III binary X-ray luminosity per unit comoving volume is
estimated to be $\log I_{\rm X} \simeq (\epsilon_{\rm bin}/0.5)[40.87 -
0.12(1+z)]$ in units of erg s$^{-1}$ Mpc$^{-3}$, which is likely
higher than those from other X-ray sources of AGNs and normal XRBs at
redshifts higher than 6 \citep{Fragos13}.  This linear fit is only
valid at $z \ge 10$ and has a maximum error of 0.5\% compared to the
computed values.  This is based on our assumption at 50\% of all Pop
III stars form in binaries, and evolve through a high mass XRB phase.
Since it takes a long time for the X-ray background to build up, the
contribution from Pop III binaries may continue to dominate the X-ray
background down to much lower redshifts.

Recently \citet{Fialkov16} used the unresolved soft X-ray background to place constraints on the nature of high redshift X-ray sources. They derived constraints on the X-ray efficiency parameter $f_x$ for three source types and two reionization scenarios, where $f_x$ is defined implicitly by the relation \citep{Furlanetto06,Fragos13} 

\begin{equation}
\frac{L_X}{SFR} = 3\times10^{40} f_X~ \rm{erg\,s}^{-1}
  \rm{M}_{\odot}^{-1}yr\ ,
\end{equation}
where here $SFR$ refers to the normal (Pop II) star formation rate density. Using the Method 2 value of $L_X \approx 5 \times 10^{39} \mathrm{erg s}^{-1} \mathrm{Mpc}^{-3}$ at $z=7.6$, and the observed SFR of $10^{-2} M_{\odot} \mathrm{yr}^{-1} \mathrm{Mpc}^{-3}$ \citep{Bouwens15}, we get $f_x \approx 16$, well within observational constraints for all cases considered.

It is interesting to compare the relative importance of ionizing UV and X-ray photons during the course of the simulations.
To $z=7.6$, both Pop III and metal-enriched stars produce about 1.4 $\times$ 10$^{71}$ erg in ionizing UV photons (or about 6.3 $\times$ 10$^{69}$ photons) 
inside the refined region, of which less than 1\% (10$^{69}$ erg), is produced by Pop III stars.  There are 
about 7.6 $\times$ 10$^{68}$ hydrogen atoms in the survey volume of the \textit{Void} run. At $z=7.6$, only $\sim$ 13.2\% hydrogen 
atoms have been ionized by the UV ionizing photons in the simulation \citep{Xu16_fesc}. So most of the ionizing UV photons are absorbed by hydrogen atoms that later
recombine and are consumed to maintain the ionization state of high density regions. On the contrary, a large fraction of X-rays are expected to 
penetrate to low density regions which have recombination times longer than the Hubble time. During the same period, the Pop III binaries 
of the \textit{Void} simulation produced $\sim$ 6.8 $\times$ 10$^{69}$ eV of energy in X-rays inside the refined region. It is about 5\% of 
energy of ionizing UV photons and about 9 eV per hydrogen atom. Using Method 2, we estimate that Pop III binaries produce in total 
2.1 $\times$ 10$^{72}$ eV of energy in X-rays in the entire simulation box. As there are about 3.6 $\times$ 10$^{71}$ hydrogen atoms in the 
simulation box, this amounts to about 6 eV of X-ray energy per hydrogen atom.  

This photon budget analysis shows that the X-ray component is not negligible. There is enough energy in X-rays from Pop III binaries to significantly change the thermal 
and ionization states of the IGM. It is very hard to know the ratio of X-ray energy used in ionizing and heating the IGM without knowing 
the SED of the X-ray sources and doing a detailed simulation \citep{Shull85}. \citet{Xu14} used a one-zone model to calculate heating 
and ionizing effects for fixed X-ray photon energies and fluxes and to estimate the temperature and  hydrogen ionization fraction of 
the \textit{Rarepeak} simulation volume by simply assuming the X-ray background is constant since z $=$ 15. This predicted that the IGM 
may be heated to several hundred Kelvin and be ionized to about $0.5\%$ by $z=6$. We have shown that the X-ray luminosity and background 
actually increase until the end of the simulation at $z=7.6$ and are at much higher ($\sim$ 10 times) levels. The electron fraction due to X-ray 
ionization might therefore reach a few percent by simply extrapolating the results in \citet{Xu14}, or even more since X-rays are also efficient at 
helium ionization. This physics is important to the understanding the optical depth of CMB. In addition, by the end of reionization, the IGM may have been homogeneously 
heated by the keV photons of the X-ray background for hundreds of Myr. The additional heat input above and beyond the UV photoheating may reach thousands of Kelvin \citep{Xu14}, and that is enough to have 
important impacts on high redshift 21-cm signatures \citep{Ahn14} and the Ly$\alpha$ forest spectra \citep{Tytler09} at lower redshifts. 
  

\acknowledgements

This research is part of the Blue Waters project, which a joint effort
supported by the NSF (award number ACI-1238993) and the state of
Illinois, using NSF PRAC OCI-0832662.  This research was supported by
NSF grants PHY-0941373, AST-1109243, AST-1211626, and AST-1333360, and
NASA grants NNX12AC98G, NNX15AP39G, HST-AR-13261.01-A,
HST-AR-13895.001, HST-AR-14315.001-A, HST-AR-14326.001.  
K.A. was supported by NRF-2012K1A3A7A03049606 and NRF-2014R1A1A2059811.
This work was performed using the open-source {\sc
Enzo} and {\sc yt} codes, which are the products of collaborative
efforts of many independent scientists from institutions around the
world. Their commitment to open science has helped make this work
possible.


\bibliographystyle{apj}

\end{document}